# Aharanov-Bohm excitons at elevated temperatures in type-II ZnTe/ZnSe quantum dots

I. R. Sellers[1*], V. R. Whiteside[1], I. L. Kuskovsky[2], A. O. Govorov[3], B. D. McCombe[1].

[1]*Dept. of Physics, Fronzcak Hall, University at Buffalo SUNY, Buffalo, NY 14260, USA*

[2]*Dept. of Physics, Queens College of CUNY, Flushing, NY 10031, USA.*

[3]*Department of Physics & Astronomy, Ohio University, Athens, OH 45701. USA*

Optical emission from type-II ZnTe/ZnSe quantum dots demonstrates large and persistent oscillations in *both* the peak energy *and* intensity indicating the formation of coherently rotating states. Furthermore, the Aharanov-Bohm (AB) effect is shown to be remarkably robust and persists until 180K. This is at least one order of magnitude greater than the typical temperatures in lithographically defined rings. To our knowledge this is the highest temperature at which the AB effect has been observed in semiconductor structures.

**[*]isellers@buffalo.edu**

The Aharanov-Bohm (AB) effect results from the motion of a charged particle as it orbits a magnetic field in a closed trajectory [1]. The phase shift developed in the envelope function of such a charged particle depends upon its sense of motion in the plane normal to the applied magnetic field and the area of its orbit (the enclosed flux). Due to recent progress in epitaxial growth and device fabrication techniques the AB effect has now been demonstrated in both metallic and semiconductor ring geometries in magneto-transport measurements [2-4], and manifests itself in oscillations in the conductance due to constructive and destructive interference of the charge carrier wavefuctions. In parallel to intensive transport studies of AB effect in lithographically-defined rings, AB oscillations were also studied in self-assembled quantum rings (QRs) by far-infrared and capacitance methods [5], and very recently by magnetization spectroscopy [6]. The experimental demonstration of the AB effect in nanostructures has lead to proposals for its implementation in quantum computation architectures [7, 8]

Since the observation of such effects in transport measurements, an interesting question has arisen concerning the possibility of observing the AB effect in neutral composite particles such as excitons, and in particular, if it is possible to observe AB oscillations in the optical properties of ring-like nanostructures. The AB effect is a property of charged particles and, in a naive picture, neutral excitons would not exhibit AB oscillations. Nevertheless, there have been several theoretical predictions [9-12] concerning a mechanism leading to AB oscillations for neutral composite quasi-particles. This is due to the fact that electron and hole in a ring-like nanostructure may move over different trajectories resulting in a non-zero electric dipole moment and, therefore, to measurable AB effects. Type-II quantum dots (QDs) have been predicted to be particularly amenable to exhibiting such AB effects because of enhanced polarization of the composite particle due to the spatial separation of the electron and hole in such systems [9-12].

Furthermore, the AB effect in QRs/QDs is predicted to have significant effects upon the luminescence properties of the nanostructures since, due to the cylindrical symmetry of the carriers, in the simplest picture, the exciton ground state has zero orbital angular momentum projection ($L=0$) at zero magnetic field, and this changes to states of higher orbital angular momentum (projection) ($L= \pm 1, 2, 3$) with increasing magnetic field. This impacts the optical properties in two ways: 1. The ground state energy oscillates as the orbital angular momentum states cross; and 2. The intensity changes from strong (bright excitonic transition with zero angular momentum projection) to weak (dark excitonic transitions for states with non-zero angular momentum projections) with increasing magnetic field [9, 12]. Until now the AB effects predicted to be observable in the luminescence have not been seen in both the energy *and* intensity in a single QD/QR structure. Although AB oscillations have been demonstrated in the emission energy in charged In(Ga)As QRs [13] and type-II InP/GaAs QDs [14], analogous effects were not observed in the PL intensity. In a very recent publication, a single AB intensity oscillation was demonstrated [15] for samples similar to those studied in this work, but oscillations in the luminescence energy were not reported.

In the present work we demonstrate strong AB oscillations in both the emission energy and intensity simultaneously from the same structure. In addition, we investigate the temperature dependence of PL and show that the AB oscillations are remarkably robust against temperature, with the signature of the AB effect visible up to 180K. This is much higher temperature than reported for lithographically-defined structures. Even in the smallest lithographically-defined rings with dimensions of few hundreds nms, the temperature at which the AB effect disappears is about 10K [4]. In general, the AB effect in a QR exists if the QR length is shorter than, or comparable to the phase-coherence length. In our type-II self-assembled QDs, the AB effect survives at elevated temperature because of the small closed trajectory of the electron in the bound exciton.

The sample studied is a 240 layer ZnTe/ZnSe superlattice structure grown by molecular beam epitaxy. Although the growth details are described elsewhere [16] we briefly detail the important points here, since the formation of the quantum dots critically depends upon the growth technique. Initially the GaAs substrate is planarized in a III-V system and then transferred under ultra-high vacuum to a II-VI reactor where the temperature is lowered to 250ºC for the deposition of a 10ML ZnSe buffer layer. This process is followed by the deposition of a Zn-Se-Te multilayer by migration-enhanced epitaxy. Initially, a ZnSe buffer layer of ~2.4nm is deposited followed by the Zn-Te-Zn active layer. Preferential migration results in clustering of Te-atoms, and eventually the formation of ZnSeTe quantum dots. The resulting 0.7nm quantum dot containing layer is then capped by a 2.4nm ZnSe barrier. This process is repeated 240 times. Recent structural studies indicated a Te-composition of at least 50% in the active region and 3-5% in the barriers [16, 17].

To understand the origin of QD formation in this multilayer superlattice structure it is necessary to describe how certain characteristic features of the photoluminescence (PL) are related to the structure of the sample. The PL at 4K excited with a HeCd laser at 3.81eV is presented in Figure 1. The observed PL displays the well known two band behavior of the ZnSeTe system [18-20]. The shoulder around 2.6 – 2.75eV (A) is generally accepted to be the result of recombination from excitons bound to pairs of Te-atoms [16, 20], which substitute isoelectronically for Se-atoms along specific crystal directions [21]. The feature at ~2.79eV (D) is related to free excitons in the alloyed Zn(Te)Se matrix [16], while the dominant feature around 2.5eV (B) has been historically attributed to the contribution of clusters of larger numbers of these isoelectronic centers. However, it was shown that these Te-clusters evolve into type-II QDs in ZnTe/ZnSe multilayer [22] structures, and even single layers, if the ZnTe layer is greater than 3 MLs thick [23]. Interestingly, a further weak feature is also evident in the PL at 2.35eV. This feature is more apparent in the inset of Figure 1, which shows the PL for the ZnTe/ZnSe structure at 180K. This low energy shoulder (C) at 2.3eV is consistent with the emission energy of *conventional* ZnTe/ZnSe self-assembled QDs grown by strain-driven processes [25], and is related to the formation of a lower density of pure ZnTe QDs.

Quantum dot formation in the sample described in this article was confirmed by time-resolved PL [22 24], which showed a clear transition from direct excitonic recombination emission to that of type-II QDs with an order of magnitude increase in the PL lifetime. The increased PL lifetime of the type-II QDs is directly related to the spatial separation of the charged particles and therefore reduction of the oscillator strength of the excitons. Furthermore, recent evidence also suggests that the QDs, which evolve from the clustering of Te-atoms in the ZnTe/ZnSe multilayer structure described here, also align vertically and therefore form columnar QD-like nanostructures [15]. This configuration is shown schematically in the inset to Figure 3(b). In such structures the electron is not only bound to the hole via coulomb attraction but is 'forced' to align laterally beside the confined hole, creating a particularly suitable geometry for the observation of AB interference since the motion is forced to be in-plane by the geometry [9, 12].

Figure 2 shows a schematic model of the orbital motion of particles inside such a localized exciton. In type-II Zn(Se)Te QDs, the hole is strongly confined by the nanostructure potential. The stacked *columnar* geometry of the system is of particular importance for the motion of the electron; it dictates the ring-like geometry of the electron motion. As was demonstrated theoretically in Ref [15], the electron inside an exciton in a single QD is localized either above or below a type-II QD and no AB effect is expected. In the stacked QDs, however, the motion is different because of the constraints due to the columnar nature of the stack: here the electron is forced to move in a plane laterally around a stack and, in this way, the AB effect can appear. In addition, numerical calculations [15] show that the tunnel coupling between QDs in a stack is very weak and an exciton cannot move in the z-direction. Therefore, we assume that an exciton is localized in one of the stacked QDs. Due to their radial spatial separation the electron and hole form a rotating dipole with the two charged particles circumscribing orbits of different area. In comparison we consider a simplified qualitative model, a rotating diapole in a magnetic field normal to the plane; the spectrum is given by [12]

$$E_{exc} = E_g + \frac{\hbar^2}{2MR_0^2}\left(L + \frac{\Delta\Phi}{\Phi_0}\right)^2, \qquad (1)$$

where $L$ is the angular momentum quantum number, $E_g$ is the confined hole to electron ground state energy, $\Phi_0 = hc/|e|$ is the magnetic flux quantum, $R_0 = (R_e + R_h)/2$, and $M = (m_e R_e^2 + m_h R_h^2)/R_0^2$; $R_h$ and $R_e$ are the averaged radii of orbits of the hole and electron respectively, and $m_h$ and $m_e$ are their masses. The quantity $\Delta\Phi = p(R_e^2 - R_h^2)B = 2p\Delta R \cdot R_0 B$ is the magnetic flux through the area between electron and hole trajectories (see Fig.2). The parameter $\Delta\Phi/\Phi_0$ can be regarded as the AB phase. If we now assume that the hole is strongly localized inside the QD (i.e. $R_h = 0$), the average distance between electron and hole becomes equal to the radius of the electron orbit, $\Delta R = R_e - R_h = R_e$ According to Eq.1, the rotating dipole shows a set of transitions between ground states with different $L$: $L = 0 \to L = -1$, $L = -1 \to L = -2$, etc. Correspondingly, the energy of the lowest exciton state as a function of the magnetic field oscillates with the period $dB = \Phi_0 / pR_e^2$. Taking the period of oscillations from our observations (2.36 T), we estimate the average radius of the electron orbit $R_e$ to be 23.5 nm. This value is in agreement with morphological studies of ZnTe/ZnSe QDs [15]. We note that the electronic radius should be somewhat larger than the structural radius of the QD because the lateral localization of the electron is mostly due to the relatively-weak Coulomb potential [15].

An interesting question is whether the e-h Coulomb interaction can affect the motion of the hole (see Fig. 2). If the Coulomb attraction is strong enough, the exciton motion should be viewed as an electron and hole rotating together. Then, the important AB parameter $\Delta\Phi/\Phi_0$ in the spectrum (1) should include a non-zero value for the orbital radius of the hole, $R_h$ (see Fig. 2). Using the numerical results from Ref. [15], we can see that the hole localization energy and the Coulomb interaction in ZnTe/ZnSe QDs are of the same order of magnitude (both about 10meV). Therefore, such correlated motion of carriers may play a role. Clearly, further theoretical and experimental investigations are necessary. Further theoretical studies would incorporate the correlated e-h motion, which was omitted in Ref. [15], and possibly strain effects. Experimentally, single QD spectroscopy is highly desirable and may add considerably to the understanding of excitonic AB oscillations and their T-dependence.

To determine the effects of magnetic field on the spectra the sample was mounted in the Faraday geometry in a 10T Oxford Instruments optical access superconducting magnet system. The magneto-photoluminescence (MPL) was measured in 0.2T steps, and the energy and intensity of the three features (isoelectronic centers (A), the ZnSeTe QDs (B) and ZnTe QDs (C)) were evaluated at each field through careful Gaussian fitting of the individual peaks. Below we focus on the intensity and position of the peak B (ZnSeTe QDs), since this peak dominates the spectrum.

The effect of magnetic field upon the energy of the type-II ZnSeTe QDs (feature B of Fig. 1) is shown in Figure 3(a). Here three well-resolved oscillations in the energy of the peak are evident, which are undamped over the magnetic field range studied. The effect of the magnetic field penetrating the ring upon the intensity of this peak is shown in Fig. 3(b). Again, clear oscillations are evident in the PL intensity, although the quality of these features is reduced. Oscillations in the luminescence intensity due to coherent AB effects in QRs/QDs were first predicted by Govorov *et al* [12], who have shown that under the condition that the electron and hole are not strongly correlated (single particle picture) the simple picture described above is modified, and a periodic switching between a ground state transition with $L=0$, to that with *L ? 0* will result with increasing magnetic field. The oscillations in the intensity of the PL shown in Fig. 3(b) may indicate such behavior for the weakly bound electron and hole in type-II ZnSeTe QDs described here. However, the selection rules for transitions in total angular momentum are only strictly valid in the situation of perfect rotational symmetry. Here it is highly probable that the ZnSeTe QDs are elongated, therefore the selection rules may be relaxed allowing the observation of *L ? 0* higher order states.

The effects of temperature are shown in Figure 4. Figures 4(a) and (b) show the magnetic field dependence of energy and intensity, respectively, of the PL for the ZnSeTe QDs (B) at 60 K. With increasing temperature, the oscillations become less visible due to de-coherence and fluctuations in the system. Nevertheless, as the inset to Figure 4(a) shows, the AB effect is evident even at 180K. This is likely to be due to the small closed trajectory of the electron. In transport measurements of the electron mobility in semiconductor nanostructures, optical phonons begin to play an important role at T > 70 K. Interestingly, the AB effect in our samples exists at temperatures higher than 70 K suggesting that the phase coherence length of the electron in the exciton persists longer than the ring length $2pR_e$. At ~100K the PL intensity, shown inset in Fig. 4(b), decreases rapidly. This is related to thermal ionization; consistent with the exciton binding energy of 7.3 meV for these QDs [22].

In summary, we have observed unusually strong Aharanov-Bohm effects in both the intensity and energy of the photoluminescence of type-II ZnSeTe quantum dots. Furthermore, temperature studies reveal that the ABE in these type-II QDs is remarkably robust, persisting at temperatures up to 180K. Although the subtleties of this behavior are not fully understood, the demonstration of quantum interference effects at high temperature indicates that type-II QDs systems may be of considerable interest for applications in quantum computation [7, 8].

We acknowledge support from DOE Award No. DEFG02-05ER46219, BSC-CUNY Award No. 68075-00 37 (ILK), and the Center for Spin Effects and Quantum Information in Nanostructures (UB). One of the authors (A.O.G.) acknowledges support by the BioNano Technology Initiative at Ohio U.

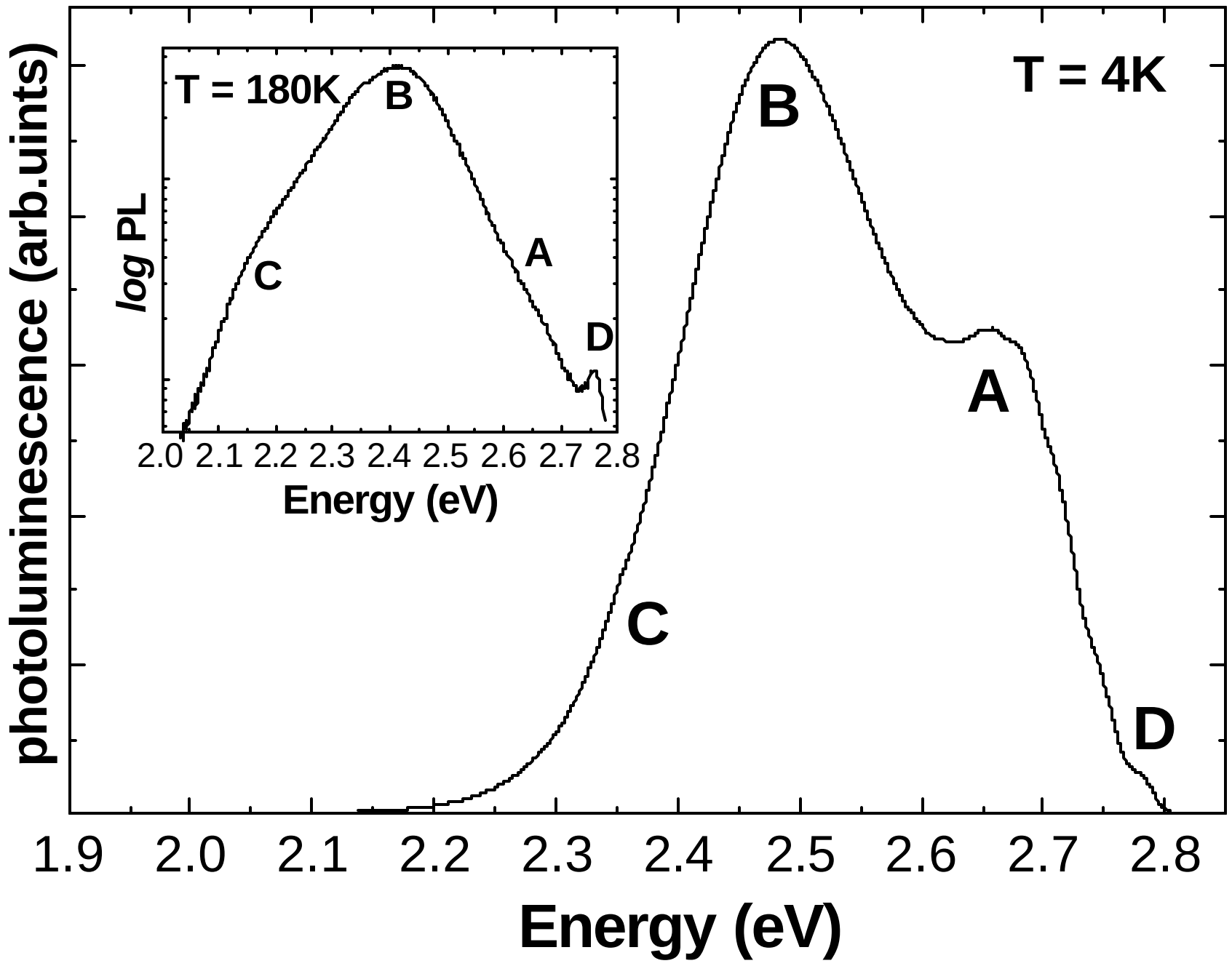

**Figure 1. Photoluminescence of the 240 period ZnTe/ZnSe superlattice structure at 4K. The contribution of isoelectronic centers (A), ZnSeTe quantum dots (B) and Te-rich ZnTe quantum dots (C) are labeled. The emission from free excitons in the Zn(Te)Se matrix is also evident at 2.79ev (D). The inset shows photoluminescence at 180K where the feature attributed to ZnTe quantum dots (C) is significant due to thermal redistribution of carriers to the low energy dots.**

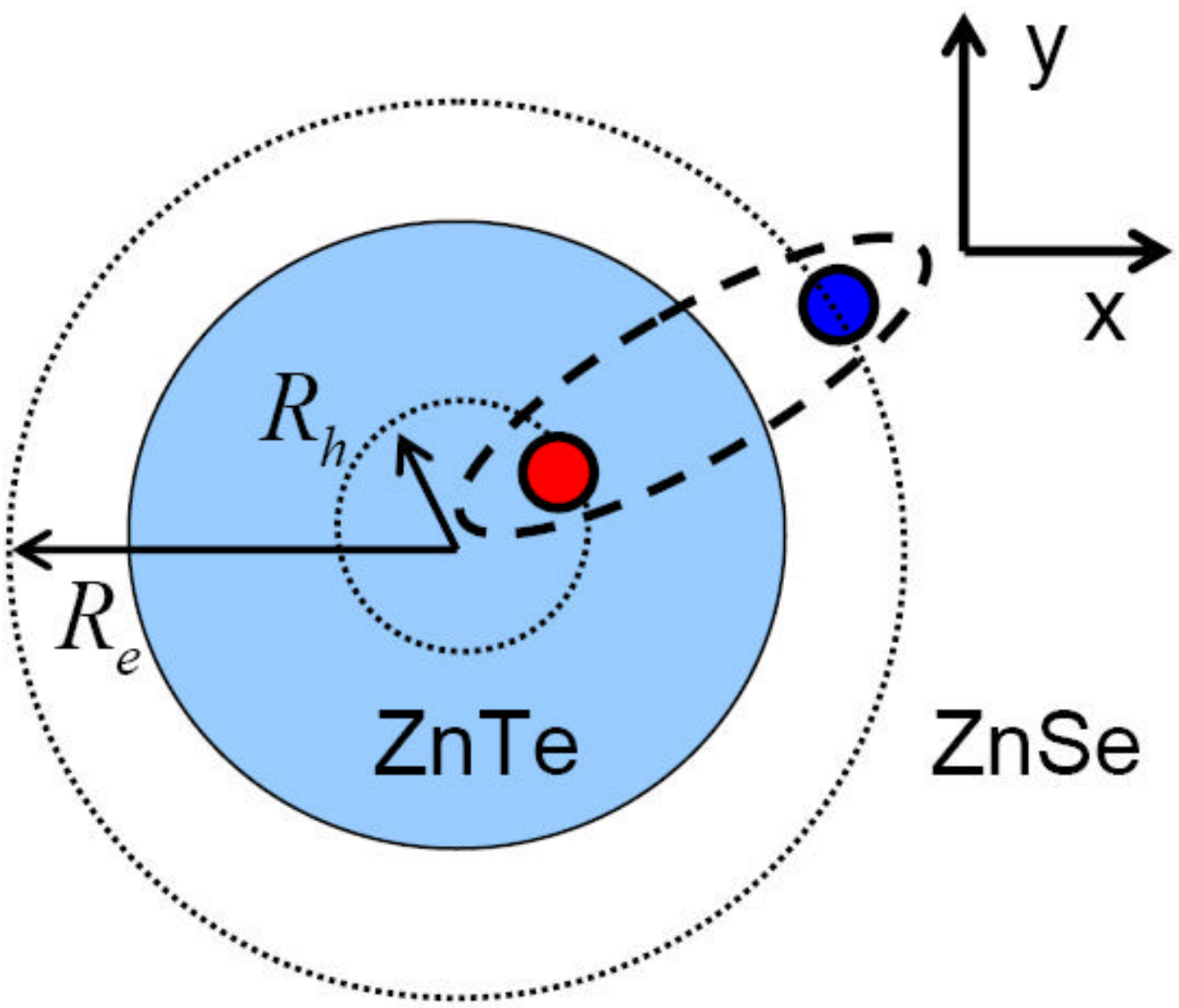

**Figure 2. Schematic of the orbital motion of particles inside a localized exciton. The x- and y-axes belong to the cross-sectional plane. The electron and hole are bound due to the Coulomb interaction. Their trajectories are shown as dotted lines.**

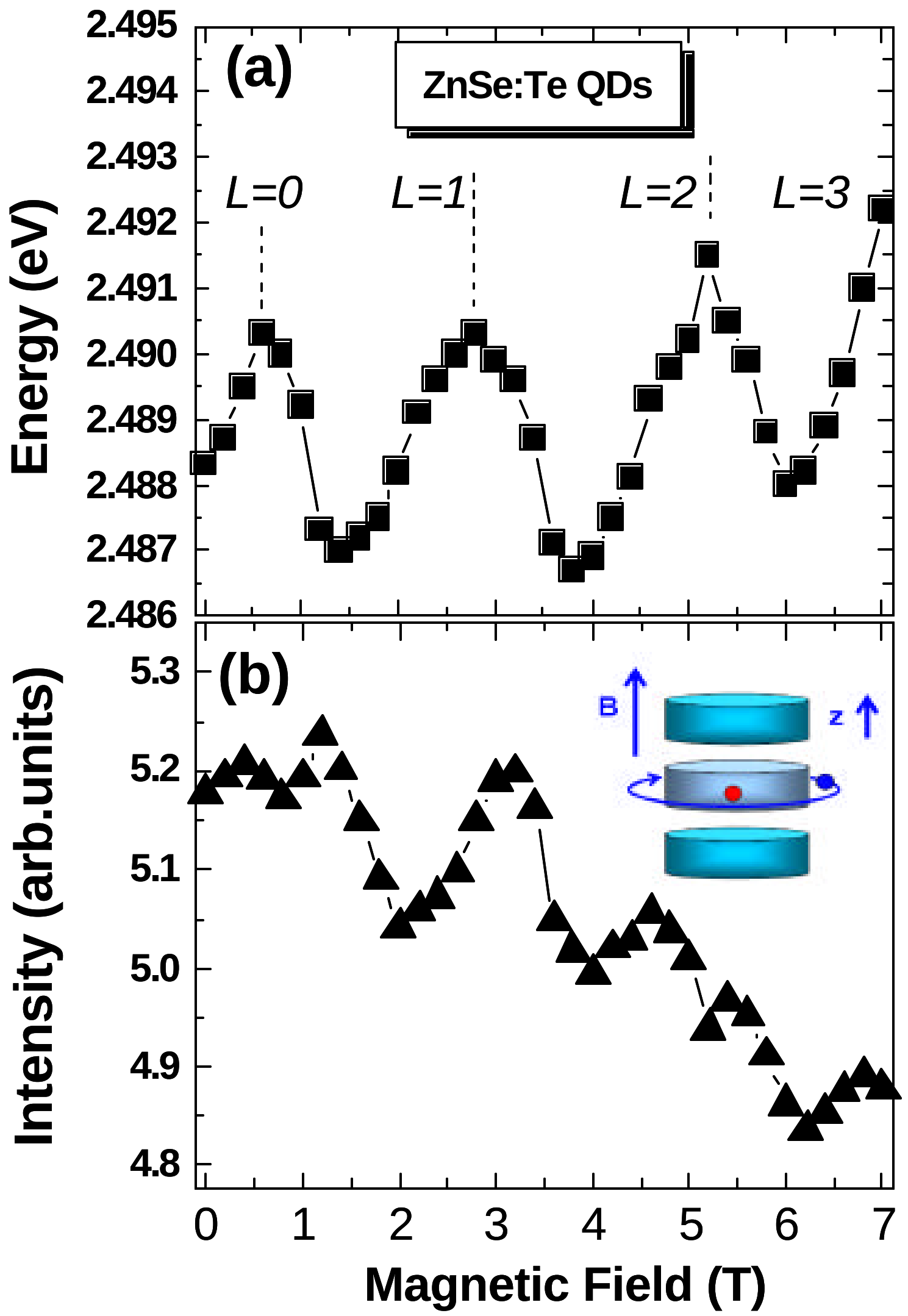

**Figure 3. Magneto-photoluminescence energy (a) and intensity (b) at 4K for the ZnSeTe quantum dots (B).** ***L=0, 1, 2 and 3*** **represent the transitions in total orbital momentum. Inset to Figure 3(b) shows a schematic representation of the columnar QD geometry.**

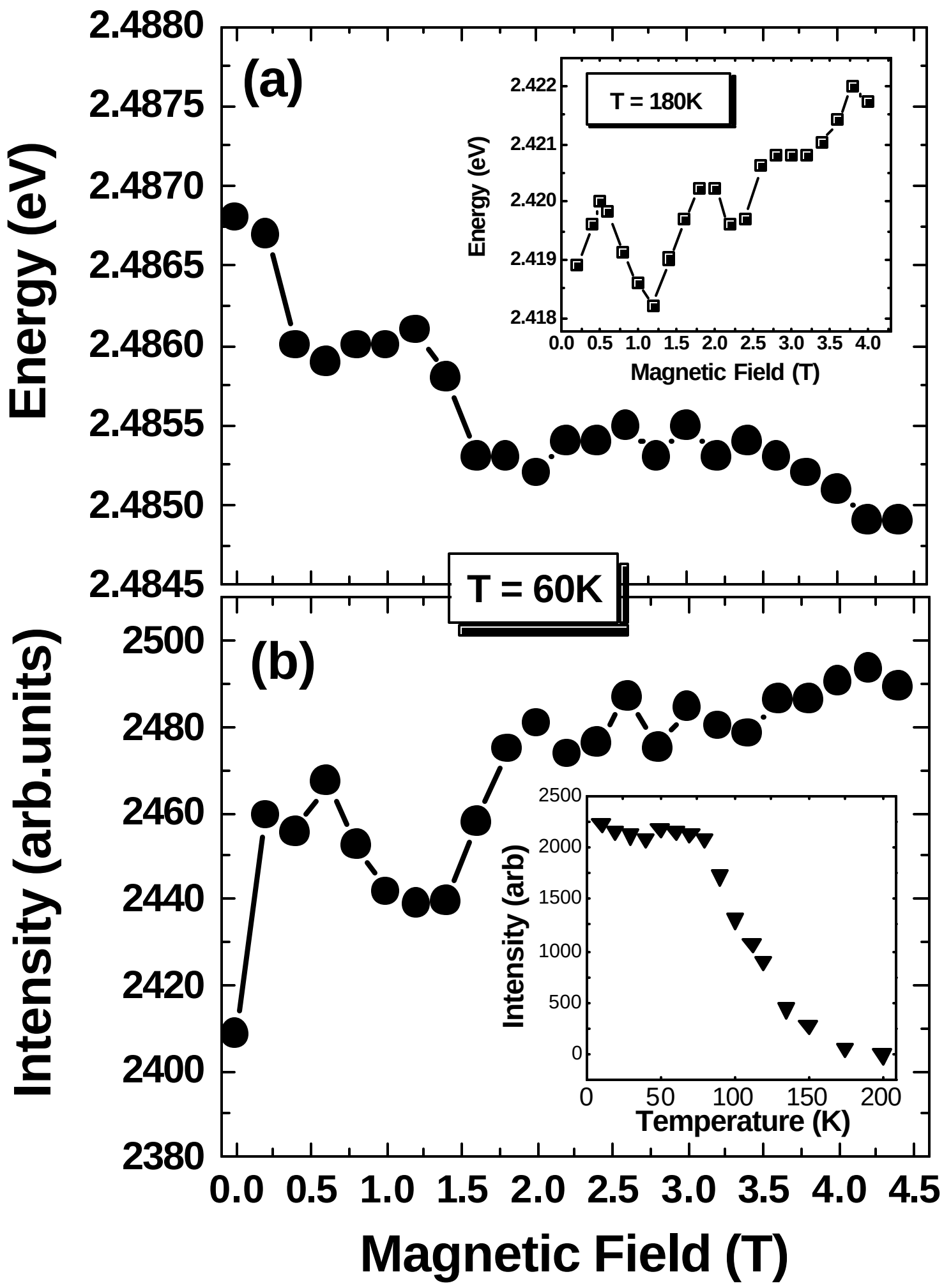

**Figure 4. Magneto-photoluminescence versus energy (a) and intensity (b) at 60K for the ZnSeTe quantum dots. The inset to Figure 4(a) shows the energy dependence at 180K while the inset to Figure 4(b) shows the temperature dependence of the integrated intensity of the dots.**